# Unveiling the orbital-selective electronic band reconstruction through the structural phase transition in $TaTe_2$

Natsuki Mitsuishi[1,2*], Yusuke Sugita[2], Tomoki Akiba[2], Yuki Takahashi[2], Masato Sakano[2,3], Koji Horiba[4], Hiroshi Kumigashira[4], Hidefumi Takahashi[5,6], Shintaro Ishiwata[5,6], Yukitoshi Motome[2], Kyoko Ishizaka[1,2,3**]

1. *RIKEN Center of Emergent Matter Science (CEMS), Wako, Saitama 351-0198, Japan.*
2. *Department of Applied Physics, The University of Tokyo, Hongo, Tokyo 113-8656, Japan.*
3. *Quantum-Phase Electronics Center (QPEC), The University of Tokyo, Hongo, Tokyo 113-8656, Japan.*
4. *Condensed Matter Research Center and Photon Factory, Institute of Materials Structure Science, High Energy Accelerator Research Organization (KEK), Tsukuba, Ibaraki 305-0801, Japan.*
5. *Division of Materials Physics and Center for Spintronics Research Network (CSRN), Graduate School of Engineering Science, Osaka University, Toyonaka, Osaka 560-8531, Japan.*
6. *Spintronics Research Network Division, Institute for Open and Transdisciplinary Research Initiatives, Osaka University, Suita, Osaka, 565-0871, Japan.*

* e-mail: natsuki.mitsuishi[at]riken.jp

** e-mail: ishizaka[at]ap.t.u-tokyo.ac.jp

**Abstract**

Tantalum ditelluride $TaTe_2$ belongs to the family of layered transition metal dichalcogenides but exhibits a unique structural phase transition at around 170 K that accompanies the rearrangement of the Ta atomic network from a "ribbon chain" to a "butterfly-like" pattern. While multiple mechanisms including Fermi surface nesting and chemical bonding instabilities have been intensively discussed, the origin of this transition remains elusive. Here we investigate the electronic structure of single-crystalline $TaTe_2$ with a particular focus on its modifications through the phase transition, by employing core-level and angle-resolved photoemission spectroscopy combined with first-principles calculations. Temperature-dependent core-level spectroscopy demonstrates a splitting of the Ta 4*f* core-level spectra through the phase transition indicative of the Ta-dominated electronic state reconstruction. Low-energy electronic state measurements further reveal an unusual kink-like band reconstruction occurring at the Brillouin zone boundary, which cannot be explained by Fermi surface nesting or band folding effects. On the basis of the orbital-projected band calculations, this band reconstruction is mainly attributed to the modifications of specific Ta 5*d* states, namely the $d_{XY}$ orbitals (the ones elongating along the ribbon chains) at the center Ta sites of the ribbon chains. The present results highlight the strong orbital-dependent electronic state reconstruction through the phase transition in this system and provide fundamental insights towards understanding complex electron-lattice-bond coupled phenomena.

**Main text**

## I. Introduction

Layered transition metal dichalcogenides (TMDCs) $MX_2$ ($M$ = transition metal, $X$ = S, Se, Te) are quasi-two-dimensional systems that exhibit a wide variety of collective quantum phenomena [1,2]. One prime example is the charge density wave (CDW) order, which is described as a long-range periodic modulation of the electronic distribution with a concomitant lattice distortion [3–5]. Many types of CDW phases have been established in TMDCs (e.g., $TaS_2$ and $NbSe_2$) with their associated electronic properties such as metal-insulator transition and superconductivity [6,7]. On the other hand, the superperiodic lattice ordering (or superstructure) had also been considered from the viewpoint of molecule-like bonding of relevant orbitals [8]. Particularly in tellurides $M$Te$_2$, the importance of $M$-Te charge transfer and Te-Te overlap has been argued, which may make the picture of $M$-$M$ and/or Te-Te local bonding more plausible as compared to sulfides and selenides [9–12]. Indeed, the $CdI_2$-type $M$Te$_2$ [see Figs. 1(a) and 1(d) for the undistorted trigonal 1*T* case] hosts various superperiodic patterns/clusters indicative of molecule-like bonding, such as Mo/W zigzag chains in (Mo,W)$Te_2$ [13],

Ir dimers in $IrTe_2$ [14], and Te dimers in $AuTe_2$ [15]. Among them, group-V $M$Te$_2$ ($M$ = V, Nb, Ta) is a common system that crystallizes in a monoclinic (3×1×3) superstructure at room temperature [Fig. 1(b); hereafter referred to as 1*T*”] [16,17]. This metallic 1*T*” state is characterized by the quasi-one-dimensional *M* double zigzag (ribbon) chains, that might also be viewed as the superposition of linear *M* trimers in two directions (black dashed lines in Fig. 1(b)) [8,9]. Related to the large and anisotropic nature of lattice distortion, several unusual phenomena have been observed in $VTe_2$, such as the vanishment of topological surface states through the 1*T*-1*T*” phase transition [18] and the photoinduced generation of transverse acoustic phonons [19,20]. Thus, the clustered $M$Te$_2$ compounds show a lot of promise for emerging novel electron-lattice-bond coupled phenomena.

Recently, tantalum ditelluride ($TaTe_2$) has attracted considerable interest due to its peculiar multiple patterns of Ta superstructures [21]. $TaTe_2$ is stable in the 1*T*” configuration [Fig. 1(b)] persisting up to the highest temperature, which is different from $VTe_2$ where the 1*T*-1*T*” phase transition occurs at a certain temperature (~480 K) [22]. With lowering temperature, $TaTe_2$ undergoes a first-order structural phase transition at $T_s$ ~ 170 K from (3×1×3) 1*T*” to a low-temperature (3×3×3) LT phase [Fig. 1(c), *C*2/*m*]. This transition is accompanied by a substantial displacement of Ta atoms along the $\mathbf{b}_m$-axis (up to ~0.28 Å), resulting in the formation of the “butterfly-like” Ta clusters [21,23]. We note that this LT structure is unique to $TaTe_2$ and has not been observed in other TMDCs including its sister compounds such as Ta$X_2$ ($X$ =S, Se) or $M$Te$_2$ ($M$ = V, Nb). At $T_s$, the electrical resistivity (magnetic susceptibility) shows an abrupt drop (increase) on cooling [21] whereas the Seebeck coefficient indicates a sign inversion [24], thereby suggesting the modifications of the electronic band structure. Recent investigations have shown that the 1*T*”-LT phase transition can be controlled by various approaches such as chemical substitution [24–27], pressure application [28], photodoping [29], reduced dimensionality [30,31], and some of them also induce superconductivity [25,26,28]. Nevertheless, the underlying mechanism of the 1*T*”-LT transition is still controversial, and multiple scenarios have been proposed in terms of Fermi surface nesting [27,32], anisotropic electron-phonon coupling [21,27], and local chemical bonding [24,33–36]. Since comparable Ta 5*d* and Te 5*p* states are theoretically predicted to reside near the Fermi level [10,11,21,29,34], resolving the orbital-specific modifications in the electronic states through the 1*T*”-LT transition may provide important clues to fully discuss this complex phase transition and related physical properties.

In this Article, we investigate the electronic structure of single crystalline $TaTe_2$ and its modifications through the 1*T*”-LT transition by utilizing core-level and angle-resolved photoemission spectroscopy (core-level PES and ARPES) with first-principles calculations. Our temperature-dependent measurements reveal a sizeable splitting of the Ta 4*f* core-level PES spectra and an unusual kink-like band reconstruction at the Brillouin zone boundary, demonstrating the significant orbital-dependent electronic state modifications in this system.

## II. Methods

Single crystalline $TaTe_2$ were grown by the chemical vapor transport method using iodine as a transport agent. Stoichiometric mixture of Ta and Te powders and iodine were sealed in evacuated quartz tubes and reacted for about one week in a three-zone furnace. The temperatures of the source and growth zones were set to 900 and 800 °C, respectively. Details regarding the sample characterizations, including the electrical resistivity, Seebeck coefficient, and low-energy electron diffraction measurements, are provided in Supplemental Material [37].

Core-level PES measurements were performed at BL28A in Photon Factory (KEK) using a system equipped with a Scienta Omicron SES2002/DA30 electron analyzer. The photon energy and energy resolution were set to 90 eV and 30 meV, respectively. High-resolution ARPES measurements were conducted at the Department of Applied Physics, The University of Tokyo, equipped with a VUV5000 He-discharge lamp and Scienta Omicron DA30 electron analyzer. The photon energy ($h\nu$) and energy resolution were set to 21.2 eV (HeIα) and 16 meV, respectively. For all measurements, the samples were cleaved in-situ at around room temperature to obtain a fresh (0 0 1) plane and the vacuum level was kept better than $2 \times 10^{-10}$ Torr throughout the measurements. The Fermi level of the samples was referenced from the Fermi-edge spectra of polycrystalline gold electrically in contact with the samples. For the ARPES data analysis, we used an empirically obtained work function of 4.9 eV.

The electronic band structures were calculated by using OpenMX code [38]. For the virtual 1*T* phase, we adopted the local density approximation (LDA) with the Perdew-Zunger parametrization for the exchange-correlation functional in the density functional theory [39] and a $8 \times 8 \times 8$ **k**-point mesh for the calculations of the self-consistent electron density and the structural optimization. We optimized the lattice structure of the virtual 1*T* phase by relaxing the primitive translational vectors and atomic positions in a non-relativistic calculation with the convergence criterion less than 0.01 eV/Å about the interatomic forces. The optimized lattice constants were $a_t$ = 3.636 Å and $c_t$ = 6.674 Å. Using the optimized structures, we calculated the electronic band structures by a relativistic *ab initio* calculation, where the relativistic effects are included by a fully relativistic *j*-dependent pseudopotential. For the 1*T*” and LT phases, we calculated the electronic band structures by the generalized gradient approximation (GGA) using the experimentally reported structural parameters in the literature [21]. We adopted the Perdew-Burke-Ernzerhof GGA functional in density functional theory [40], a $8 \times 8 \times 8$ **k**-point mesh, and a fully relativistic *j*-dependent pseudopotential for the calculations of the self-consistent electron density. For direct comparison with ARPES data, we constructed the proper spectral weights by employing the unfolding procedure [41,42] based on the primitive (virtual) 1*T* unit cell.

## III. Results

### A. Core-level spectra.

First, we demonstrate the temperature dependence of the core-level PES spectra that provides direct insight into the atomic-site specific electronic modulations. Figure 2(a) presents the overall core-level spectra in the high-temperature 1*T*” phase (320 K) collected using a synchrotron light source ($h\nu$ = 90 eV). The intense Ta 4*f* and Te 4*d* doublet peaks are discerned at binding energies ($E_B$) of 22–25 and 39–42 eV, respectively, together with other core-levels (see the magnified profile depicted by orange). Figures 2(b) and 2(c) show the temperature dependence of the Ta $4f_{7/2}$ core-level intensity demonstrated as the color map and the set of spectral profiles, respectively. They are collected in the temperature range from 320 to 20 K during the cooling process. The spectral profile of Ta $4f_{7/2}$ above $T_s$ (~170 K) is broad and asymmetric with its peak position nearly fixed at $E_B$ ~ 22.7 eV. Upon lowering the temperature below $T_s$, the peak position gradually shifts toward lower $E_B$ and a multi-peak structure emerges. This temperature-dependent splitting is reproduced in the temperature-cycle and also by bulk-sensitive soft x-ray measurements [37], thereby substantiating its bulk origin. The fitting analysis using Voigt functions with a Sherly-type background yields the maximum splitting energies of 0.11(3) eV and 0.31(3) eV for 1*T*” (320 K) and LT (20 K), respectively, as displayed by the black markers and dotted curves in Fig. 2(c) (see Supplemental Material [37] for more details). This implies that the variation of the electron densities among the inequivalent Ta sites (i.e., *d*-*d* charge transfer) is substantially enhanced across the transition. Meanwhile, we note that these splitting energies are much smaller as compared to those in the Star-of-David CDW systems such as 1*T*-$TaS_2$ (~1.2 eV) [43] and 1*T*-$TaSe_2$ (~0.9 eV) [44], where strong electron-electron correlation effect (Mott transition) often intertwines with the $\sqrt{13} \times \sqrt{13}$ superperiodic state [6,45,46].

Similarly, the temperature dependence of the Te $4d_{5/2}$ core-level is extracted as shown in Figs. 2(d) and 2(e). Since there are three inequivalent Te sites in 1*T*”, the spectra above $T_s$ already feature a multi-peak structure. In contrast to Ta $4f_{7/2}$, the Te $4d_{5/2}$ spectral distribution hardly changes even below $T_s$. This indicates that the charge redistribution among the Te sites (*p*-*p* charge transfer) across the transition is negligible compared to that of the Ta sites. On the other hand, we find a slight shift of the peak maximum toward a lower $E_B$ at lower temperatures as indicated by the orange circle markers in Fig. 2(d). This is more clearly seen in the overlaid profiles at 320 K (red) and 20 K (blue) in Figs.

2(e) and 2(f), where the peak positions differ by about 45 meV. To examine this shift more precisely, we compare the temperature dependence of the energy shift (relative to the value at 20 K) estimated by the position of the highest intensity peak for Te $4d_{5/2}$ (orange circles) and the center of the spectral weight for Ta $4f_{7/2}$ (purple triangles) and Te $5d$ (including both $4d_{5/2}$ and $4d_{3/2}$, cyan rectangles), as displayed in Fig. 2(g). Remarkably, both the Ta $4f$ and Te $5d$ shifts show nearly the same temperature dependence, indicating that the amount of the Ta-Te charge transfer (*d*-*p* charge transfer) is not significantly modified upon the 1*T*"-LT transition. We attribute this uniform shift mainly to the temperature-dependent chemical potential shift, which is also consistent with the temperature-dependent variation of the valence band position as highlighted by the gray shaded area (see Supplemental Material for this estimation [37]). This behavior is in stark contrast to another $5d$ system $IrTe_2$, where the trigonal-triclinic transition at ~280 K is accompanied by the Ir dimerization and significant *d*-*p* charge transfer, resulting in the core-level PES splitting of both Ir $4f$ and Te $5d$ [47]. The present results of core-level PES on $TaTe_2$ thus suggest that the Ta state is strongly modified whereas the Te state remains mostly intact through the 1*T*"-LT phase transition.

### B. Electronic band structure.

Next, we focus on the valence band structure. For simplicity, we use a set of high-symmetry points based on the primitive (virtual) 1*T* unit cell (see Fig. 1(e) for the (0 0 1) surface Brillouin zone (BZ)). Here we define $\overline{\mathrm{K}}_1/\overline{\mathrm{K}}_2$ and $\overline{\mathrm{M}}_1/\overline{\mathrm{M}}_2$ points that become inequivalent in the 1*T*" phase as compared to virtual 1*T*. Figure 3(a) shows the ARPES intensity map at $E_F$ recorded in the high-temperature 1*T*" phase (300 K), obtained by using a He-discharge lamp ($h\nu = 21.2$ eV). We find the signature of multiple warped Fermi surfaces extending along the $\overline{\Gamma}-\overline{\mathrm{M}}_1$ ($k_x$) direction, which is perpendicular to the Ta ribbon chains in the real space (i.e., the $\mathbf{b}_m$-axis). We note that such highly anisotropic Fermi surfaces are successfully observed because of the well-separated single domain. In the case of multi-domain samples, the signal mixing from multiple in-plane 120-degree domains should be carefully considered [18,48,49]. Figure 3(b) displays the ARPES image at 300 K recorded along the $\overline{\Gamma}-\overline{\mathrm{K}}_{1(2)}-\overline{\mathrm{M}}_{1(2)}-\overline{\Gamma}$ lines. Of particular interest relevant to the quasi-one-dimensional Fermi surface are the bands residing at the virtual 1*T* BZ boundaries. There are several dispersive bands that cross the Fermi level ($E_F$) along $\overline{\mathrm{M}}_1-\overline{\mathrm{K}}_1$. In contrast, two relatively flat bands lie at $E_B$ ~ 0.6 and 0.9 eV along $\overline{\mathrm{M}}_2-\overline{\mathrm{K}}_2$. These characteristic bands, together with the quasi-one-dimensional Fermi surfaces, are very similar to those observed in the isostructural 1*T*"-$VTe_2$ [18] and are also qualitatively well reproduced within our band unfolding calculation ($k_z = 0$) as shown in Figs. 3(c) and 3(d).

In Figs. 3(e)–(h), we present the results for the low-temperature LT phase in a similar manner as in Figs. 3(a)–(d). The ARPES data shown in Figs. 3(e) and 3(f) are collected at 15 K. While at first glance the Fermi surface contour in Fig. 3(e) is almost unchanged from 1*T*” [Fig. 3(a)], we observe a substantial spectral reconstruction in the band dispersion [Fig. 3(f)]. Along $\bar{\Gamma}-\bar{\mathrm{K}}_2$ and $\bar{\Gamma}-\bar{\mathrm{M}}_2$, the hole-like bands split into sharp submanifolds with small energy gaps (typically 50–150 meV), as denoted by the red arrows. These features are well reproduced by our band calculation as shown in Fig. 3(h). While most of the spectral weight is concentrated on the original 1*T*” bands, a closer comparison between ARPES and calculation [37] can also trace the additional faint band structures that are distributed with the LT (3×3) periodicity. Thus, we attribute these modifications essentially to the band-folding effect [41]. On the other hand, at around $\bar{\mathrm{M}}_1$ (the BZ boundary), a linear band dispersion gets strongly modified as marked by the black arrow in Fig. 3(f), which is difficult to explain by the simple band-folding effect (will be discussed later).

To further inspect the band modifications across the transition, we systematically compare the ARPES data near $E_F$ along several momentum cuts covering the 1*T* BZ, as shown in Figs. 4(a) and 4(b). The positions of the momentum cuts #1–#7 are depicted on the Fermi surface map in Fig. 4(c). Here, the data at 300 K [Fig. 4(a)] are divided by the Fermi-Dirac distribution function convoluted with the Gaussian resolution function to remove the thermal smearing effect of the Fermi-Dirac cutoff. The cyan curves and blue circle markers in Figs. 4(a) and 4(b) indicate the momentum distribution curves at $E_F$ and their peak positions (i.e., Fermi momentum), respectively. We find several types of modifications on cooling, such as band segmentation predominantly seen in cuts #1, #2, and the upward shift of the $E_F$-crossing Λ-shaped band as depicted by the white dotted lines in cuts #3, #4. These changes can be well explained by the band-folding effect and chemical potential shift described above, respectively. Rather, the most striking change is found in the intense V-shaped inner band around the BZ boundary (cuts #6, #7), centered at $k_y = 0$. In cut #7 (i.e., $\bar{\mathrm{K}}_1-\bar{\mathrm{M}}_1-\bar{\mathrm{K}}_1$), this ‘inner’ V-shaped bands transform into a pair of less-dispersive kink-like structures with about 70% decreased gradient, and their near-$E_F$ spectral intensity becomes abruptly suppressed at an endpoint momentum $k_1$’. Indeed, as displayed in Fig. 5(d), the energy distribution curve at $k_1$’ exhibits a characteristic peak at $E_B$ ~ 70 meV (the triangle marker), which is absent in that at the Fermi momenta ($k_1$) in 300 K. Again, this band reconstruction goes beyond the simple band-folding effect. This is in line with a recently reported temperature-dependent optical conductivity spectrum [50], where the sudden decrease in the Drude width (i.e., scattering rate) is observed upon cooling across $T_s$, together with the appearance of a sharp peak at 800 cm$^{-1}$ indicative of some inter-band transition. This energy (~100 meV) is close to that of the strong band reconstruction observed in the present ARPES (~70 meV), as depicted by the black triangle markers in Figs. 4(b) and 4(d).

Here we remark on Fermi surface nesting, which has been discussed in some recent works [27,32] as the possible driving force of the 1*T*”-LT transition. In our ARPES results, while it is difficult to fully capture the Fermi surface topology due to the complex band structures even for 1*T*”, we still readily identify a set of warped Fermi surfaces around $|k_y|$ ~0.4 Å$^{-1}$, as their Fermi momenta highlighted by the blue filled markers in Figs. 4(a) and 4(c). This ‘outer’ Fermi surface is also observed in the LT phase [Fig. 4(b)] with a slight increase in $|k_y|$ seemingly due to the chemical potential shift. To examine its temperature dependence, we show in Fig. 4(d) the energy distribution curves at the Fermi momenta in cuts #5–7 (labeled as $k_2$–$k_4$ and $k_2$’–$k_4$’ for 300 and 15 K, respectively), where the ‘outer’ bands are clearly resolved. For all spectra in both phases, the spectral weight in the vicinity of $E_F$ obeys the Gaussian-convoluted Fermi-Dirac distribution function, similar to the numerical simulation for 300 K and the experimental Au spectra at 15 K shown in the bottom of Fig. 5(d). This indicates that the ‘outer’ Fermi surface is retained without any gap formation at $E_F$ through the transition and is thus irrelevant to the Fermi surface nesting. Regarding the ‘inner’ bands (located in $|k_y|$ < 0.35 Å$^{-1}$) including the V-shaped band at $\overline{\mathrm{M}}_1$, they are strongly modified at LT. Compared to the **q**-vector size of the LT (3×3) periodic lattice distortion ($|\mathbf{q}_{LT}|$ ~ 0.67 Å$^{-1}$, see Fig. 1(e)), however, the length of the momentum vector connecting these ‘inner’ bands along the $\mathbf{q}_{LT}$ direction is far short (at most ~0.4 Å$^{-1}$). Therefore, our results do not support the simple Fermi surface nesting scenario as the origin of the 1*T*”-LT transition.

**C. Orbital character.**

Now we discuss the band orbital character. We introduce the regular octahedral coordination with orthogonal *XYZ* axes [see Fig. 1(c)], where the distortions in the actual $TaTe_6$ octahedrons are omitted. This is a very simple but helpful model for capturing the band properties at the BZ boundary in the trigonal 1*T*-type TMDCs, based on the concept of the “hidden Fermi surface” raised by Whangbo *et al*. [8,51]. Here, the three equivalent $t_{2g}$ *d* orbitals ($d_{XY}/d_{YZ}/d_{ZX}$) respectively form one-dimensional σ-bonding states along the edge-sharing octahedral network, and their combination can constitute the hypothetical Fermi surface as illustrated in Fig. 5(a). Since the σ-bonding is parallel to K–M–K, the resulting band structure exhibits a dispersive V-shape along this direction [see Fig. 5(b) for the virtual 1*T*-$TaTe_2$]. Though the hybridization with the chalcogen *p*-orbitals should be carefully considered for each case [11], this concept has been essentially demonstrated by first-principles calculations for various metallic 1*T*-type TMDCs [18,52,53].

The above concept can be further extended to discuss the band characters in the present 1*T*” and LT phases of $TaTe_2$, as shown in Figs. 5(c)–(e) and 5(f)–(h), respectively. Figures 5(c)–(e) respectively show the ARPES image, band unfolding calculation, and Ta 5*d* orbital-projected band calculations along $\bar{\mathrm{K}}_1-\bar{\mathrm{M}}_1-\bar{\mathrm{K}}_1$ and $\bar{\mathrm{K}}_1-\bar{\mathrm{M}}_2-\bar{\mathrm{K}}_2$, obtained for the 1*T*" phase. The size and color of the markers in Fig. 5(e) represent the total amount of the $t_{2g}$ *d*-orbital contribution and the ratio of the *XY*/*YZ*/*ZX* component, respectively, at each eigenstate. As described above, the strongly anisotropic quasi-one-dimensional band structures are realized in 1*T*", i.e., the $E_\mathrm{F}$-crossing V-shaped band at $\bar{\mathrm{M}}_1$ and the two flat bands at $\bar{\mathrm{M}}_2$, as marked by the green and red/blue arrows in Fig. 5(c), respectively. These bands are well reproduced by our calculations [Figs. 5(d) and 5(e)] and are mainly derived from the Ta $d_{XY}$ and $d_{YZ}/d_{ZX}$ orbitals, respectively (note that the *Z* axis is perpendicular to the Ta chain direction, the $\mathbf{b}_\mathrm{m}$-axis). It can also be recognized in the calculated partial density of states (PDOS) distributions in Fig. 5(i). Under the (hypothetical) trigonal 1*T* phase, the $d_{XY}/d_{YZ}/d_{ZX}$ orbitals are equivalent as shown in the left panel of Fig. 5(i). As for the two-fold monoclinic 1*T*" [the middle panel in Fig. 5(i)], in contrast, the $d_{YZ}/d_{ZX}$ (red-blue curve) shows the sharp peaks at $E_\mathrm{B}$ ~ 0.6 and 0.8 eV corresponding to the flat bands, whereas the $d_{XY}$ (green curve) is rather featureless in this energy region. We note that a similar electronic structure has been already demonstrated in the sister compound (V,Ti)$Te_2$, where the formation of the $d_{YZ}/d_{ZX}$ flat bands on cooling are observed at $E_\mathrm{B}$ ~ 0.2 eV around the $\bar{\mathrm{M}}_2$ point [18]. There, the flat bands were discussed in terms of the localized electronic state formed by the molecular-like trimerized vanadium bonding.

Figures 5(f)–(h) show the band dispersions for the LT phase, in the similar manner as Figs. 5(c)–(e). Although the original calculated band structure for LT (the thin curves in Fig. 5(h)) contains so many branches due to the folding, its unfolded image [Fig. 5(g)] is directly comparable to the ARPES data [Fig. 5(f)] and the 1*T*" results [Figs. 5(c) and 5(d)]. Both the ARPES spectra and calculations indicate that the $d_{YZ}/d_{ZX}$-dominated flat bands at $\bar{\mathrm{M}}_2$ are scarcely modified through the 1*T*"-LT transition, whereas the $d_{XY}$-derived V-shaped band at $\bar{\mathrm{M}}_1$ exhibits a substantial reconstruction into unusual kink-like structures. To discuss this modification, we present in Fig. 5(j) the Ta site-specific PDOS calculations for $d_{XY}$ in the 1*T*" and LT phases (the full set of Ta PDOS data are provided in Supplemental Material [37]). While the PDOS for Ta1/Ta2 in 1*T*" and Ta2A/Ta2B in LT [see Figs. 1(b) and 1(c) for the site notations] exhibit the broad shape with no significant structures below $E_\mathrm{F}$, those for Ta1A/Ta1B in LT exhibit the multiple peak structures in a wide $E_\mathrm{B}$ region. The peak just below $E_\mathrm{F}$ (black arrows) partly reflects the endpoint of the kink-like band structure observed in $\bar{\mathrm{K}}_1-\bar{\mathrm{M}}_1-\bar{\mathrm{K}}_1$ (the green arrows in Figs. 5(f)–(h)). Hence, we suggest that the $d_{XY}$ orbital states of Ta1A/1B are predominantly modified through the 1*T*"-LT transition.

We add notes on the orbital character of the 'outer' Fermi surface discussed in Fig. 4. The band calculation in 1*T*" [Figs. 5(d) and 5(e)] reproduces the 'outer' band near $\bar{\mathrm{K}}_1$ as marked by the gray arrow. The corresponding marker sizes as displayed in Fig. 5(e) are fairly small, indicating its Te 5*p*-dominanted nature. The band calculations for the LT phase [Figs. 5(g) and 5(h)] show that this 'outer' band crossing $E_F$ remains with similar dispersion relation, which agrees well with the experimental observations. This robustness of the Te 5*p*-derived state through the transition is also consistent with the temperature-independent Te 4*d* core-level PES spectra presented in Figs. 2(d) and 2(e).

## IV. Discussion and Conclusion

The present findings can be summarized as follows: (i) The core-level PES showed the temperature-dependent splitting of the Ta $4f_{7/2}$ core-level spectra through the transition (up to ~0.31 eV) in stark contrast to Te $4d_{5/2}$, suggesting the electronic reconstruction occurring predominantly on Ta sites. (ii) The temperature-dependent ARPES measurements (from 300 K to 15 K) revealed notable modifications including the sizeable chemical potential shift (~45 meV), additional (3×3)-folding of band dispersions, and the unusual kink-like band reconstruction ($E_B$ ~ 70 meV) occurring around the BZ boundary ($\overline{M}_1$ point) in the LT phase. (iii) Comparison with the orbital-projected band calculations indicated that the kink-like band reconstruction was primarily originated from modifications in the $d_{XY}$ orbital states at the center Ta sites of the zigzag ribbons (i.e., Ta1 and Ta1A/1B), whereas the $d_{YZ}/d_{ZX}$-derived flat bands (around $\overline{M}_2$) and the Te 5*p*-dominated 'outer' Fermi surfaces remained relatively unaffected throughout the transition. These results highlight the strong orbital-selective electronic modifications in this material. Our results also rule out a simple Fermi surface nesting scenario as the origin of the 1*T*''-LT phase transition and may instead suggest some additional chemical bonding involving the Ta1 (Ta1A/1B) sites. Indeed, recent x-ray diffraction studies [36,54] imply the possible role of Ta-Ta chemical bonding that induces the instability of the Ta1 position. However, the mechanism that drives the transformation from the ribbon chain to the butterfly-type Ta patterns needs to be further investigated, e.g., by using the out-of-equilibrium ultrafast experiments.

Finally, we would like to discuss the inherent structural fluctuation emerging above $T_s$, that is relevant to the present ARPES results. According to the work by Sörgel *et al.* [21], the room-temperature $TaTe_2$ hosts an anomalously large atomic displacement parameter of Ta1 atoms along the $\mathbf{b}_m$-axis (i.e., $U_{22}$ ~ 0.04 Å$^2$, which exceeds more than twice the values of $U_{11}$ and $U_{33}$). This indicates that the ribbon chain pattern in $TaTe_2$ has some prominent instability towards additional distortion, appearing as the strong fluctuation of the Ta1 atomic positions. We also point out that an unusual convex-upward electrical resistivity curve [21,24,27,28,34,50,55] as well as a broad Drude component in the optical conductivity spectrum (about 20 times broader from 300 K to 5 K) [50] are also reported above $T_s$, suggesting that the electrons are strongly scattered in this regime. In this respect, our ARPES data at 300 K [Fig. 5(c)] already exhibits the faint kink-like feature near $E_F$ (marked by the green arrow), together with the unusually blurred spectral weight lying at $E_B$ ~ 0.5 eV between $\bar{M}_1$ and $\bar{K}_1$ (marked by the white arrow). Although anomalously broad, these features are rather characteristic of the LT phase [Fig. 5(f)–(h)] and are completely lacking in the 1*T*” calculation [Figs. 5(d,e)] under static atomic configurations. This indicates that the strong Ta1 atomic fluctuation indeed affects the electronic band structure, giving rise to the anomalous electronic properties. In this viewpoint, a comprehensive structure analysis on the bulk group-V $MTe_2$ ($M$ = V, Nb, Ta) would be informative to elucidate whether such structural fluctuation is ubiquitous in 1*T*”-type $MTe_2$ compounds [16,17,54].

In conclusion, we systematically investigated the electronic structure of $TaTe_2$ through the 1*T*”-LT structural phase transition by using core-level PES, ARPES, and first-principles calculations. The present results reveal the strong orbital-dependence of electronic state modifications through the 1*T*”-LT transition, as demonstrated by the Ta 4*f* core-level splitting and the kink-like band reconstruction at the BZ boundary. Our findings provide important keys towards understanding the complex electron-lattice-bond coupled phenomena emerging in various metallic systems.

**Acknowledgements**

The authors acknowledge Naoyuki Katayama for valuable discussions and Satoshi Yoshida, Bruno Kenichi Saika, Ryu Yukawa, and Miho Kitamura for soft x-ray PES measurements. N.M. and Y.S. acknowledge the support by the Program for Leading Graduate Schools of the University of Tokyo, Advanced Leading Graduate Course for Photon Science (ALPS) and Material Education program for the future leaders in Research, Industry and Technology (MERIT), respectively. Y.S. acknowledges the supports by Japan Society for the Promotion of Science through a research fellowship for young scientists. Parts of the numerical calculations were performed in the supercomputing systems in ISSP, the University of Tokyo. The experiments were partly performed under KEK-PF proposals (Nos. 2018G093, 2018G624). This work was partly supported by the JSPS KAKENHI (Nos. JP20H01834, JP21H01030, JP21H05235).

**Figures & captions**

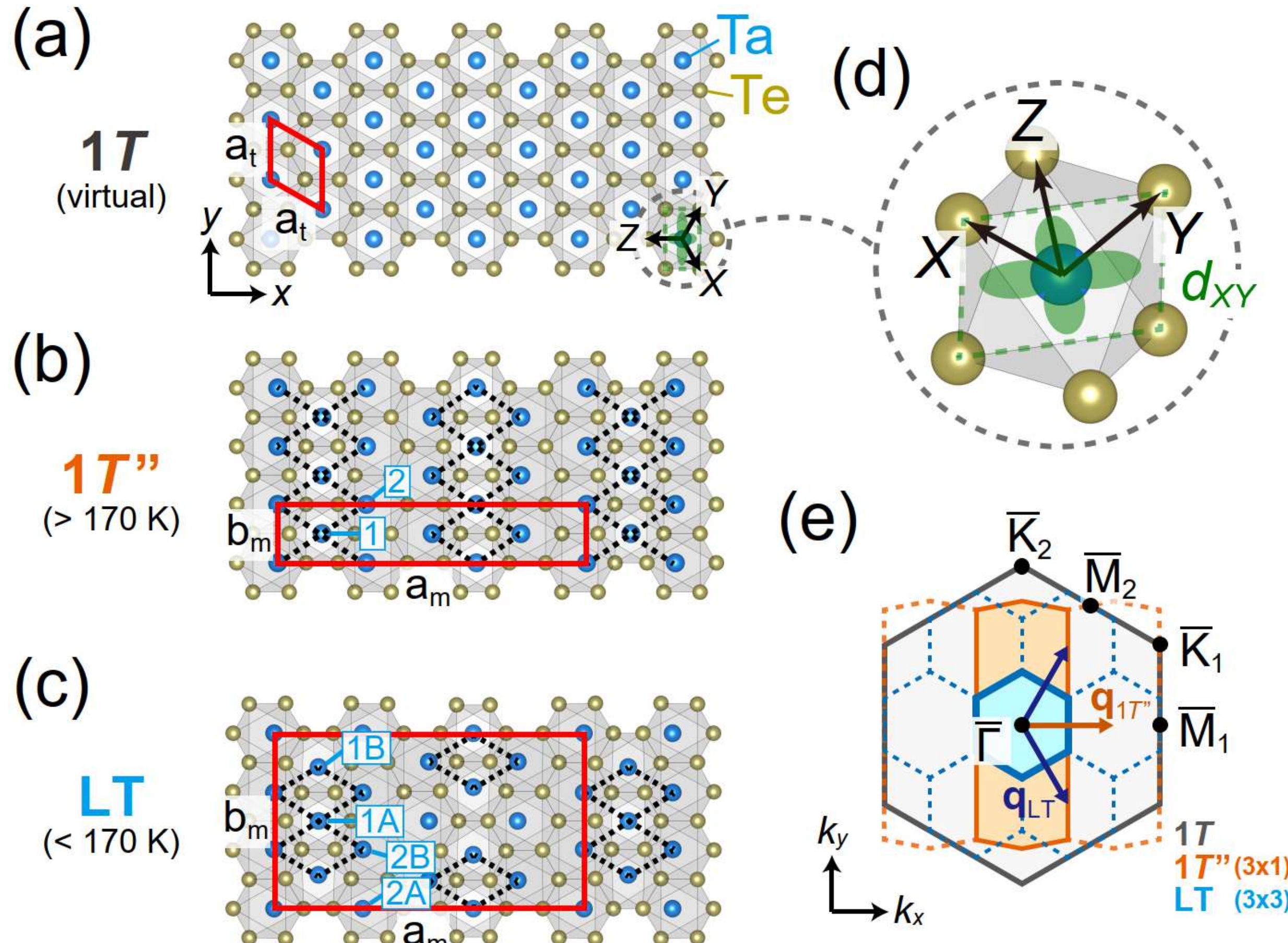

**Fig. 1. Crystal structure and (0 0 1) surface Brillouin zone of $TaTe_2$.** (a)–(c) Top views of $TaTe_2$ layer for the virtual undistorted 1*T* [(a), space group: $P\bar{3}m1$], high-temperature 1*T''* (above 170 K) [(b), $C2/m$], and low-temperature LT phases (below 170 K) [(c), $C2/m$]. The red squares indicate the (conventional) unit cells. The black broken lines in (b) and (c) highlight the characteristic Ta-Ta cluster patterns proposed in the literature [21]. The labels denote the inequivalent Ta sites. (d) $TaTe_6$ octahedron and the local orthogonal coordination (*XYZ*) adopted for the orbital-projected band calculations shown in Fig. 5. The $d_{XY}$ orbital is depicted as an example. (e) (0 0 1) surface Brillouin zones for 1*T* (gray), 1*T*'' (cyan), and LT (orange). $\mathbf{q}_{1T''}$ and $\mathbf{q}_{\mathrm{LT}}$ indicate the **q**-vector of the 1*T''* (3×1) and LT (3×3) periodic lattice distortions, respectively. The crystal structures are visualized by VESTA [56].

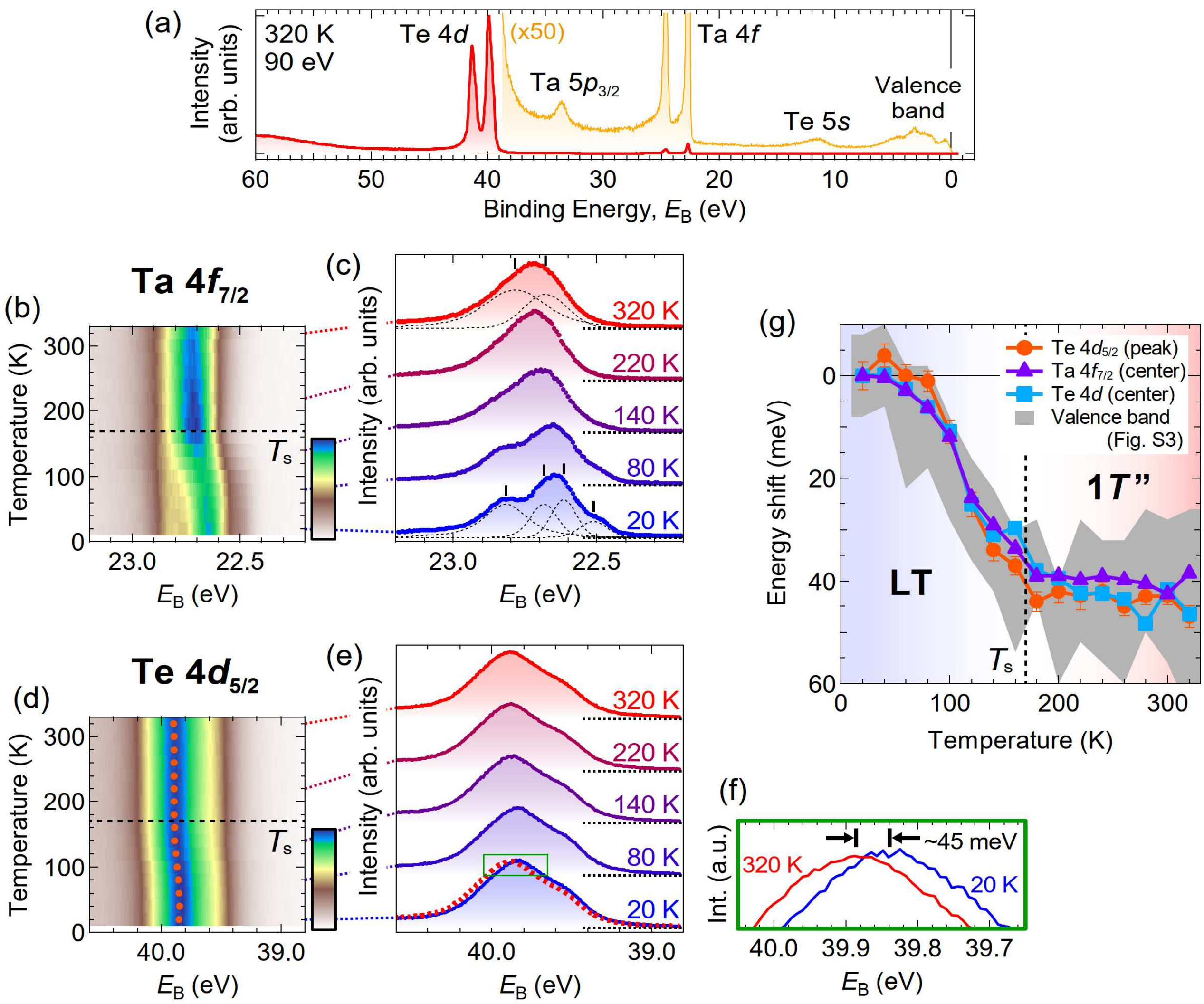

**Fig. 2. Temperature dependence of core-level spectra.** (a) Overall core-level photoemission spectra measured at 320 K. The magnified profile (×50, orange) is also shown for clarity. (b), (c) Temperature-dependent evolution of the Ta $4f_{7/2}$ core-level intensity color map [(b)] and spectral profiles at selected temperatures [(c)]. The black dotted curves in (c) indicate the fitting Voigt functions at 320 K and 20 K [37]. (d), (e) Same as (b) and (c), but for the Te $4d_{5/2}$ core-level. The orange circle markers in (d) trace the highest intensity peak positions. The spectrum at 320 K (the broken red curve) is also overlaid on the data at 20 K for comparison in (e). (f) Close-up comparison of the peak top of Te $4d_{5/2}$ spectra (the green rectangle in (e)) at 320 K and 20 K. (g) Temperature dependence of energy position shifts (relative to the values at 20 K) for the highest intensity peak of Te $4d_{5/2}$ (orange circles), the center of spectral weight of Ta $4f_{7/2}$ (purple triangles) and Te $4d$ (including $4d_{5/2}$ and $4d_{3/2}$, cyan rectangles). The gray shaded region shows the possible energy shift estimated from the temperature-dependent valence band measurements [37]. All the data in this figure were collected with a synchrotron light source (photon energy: 90 eV).

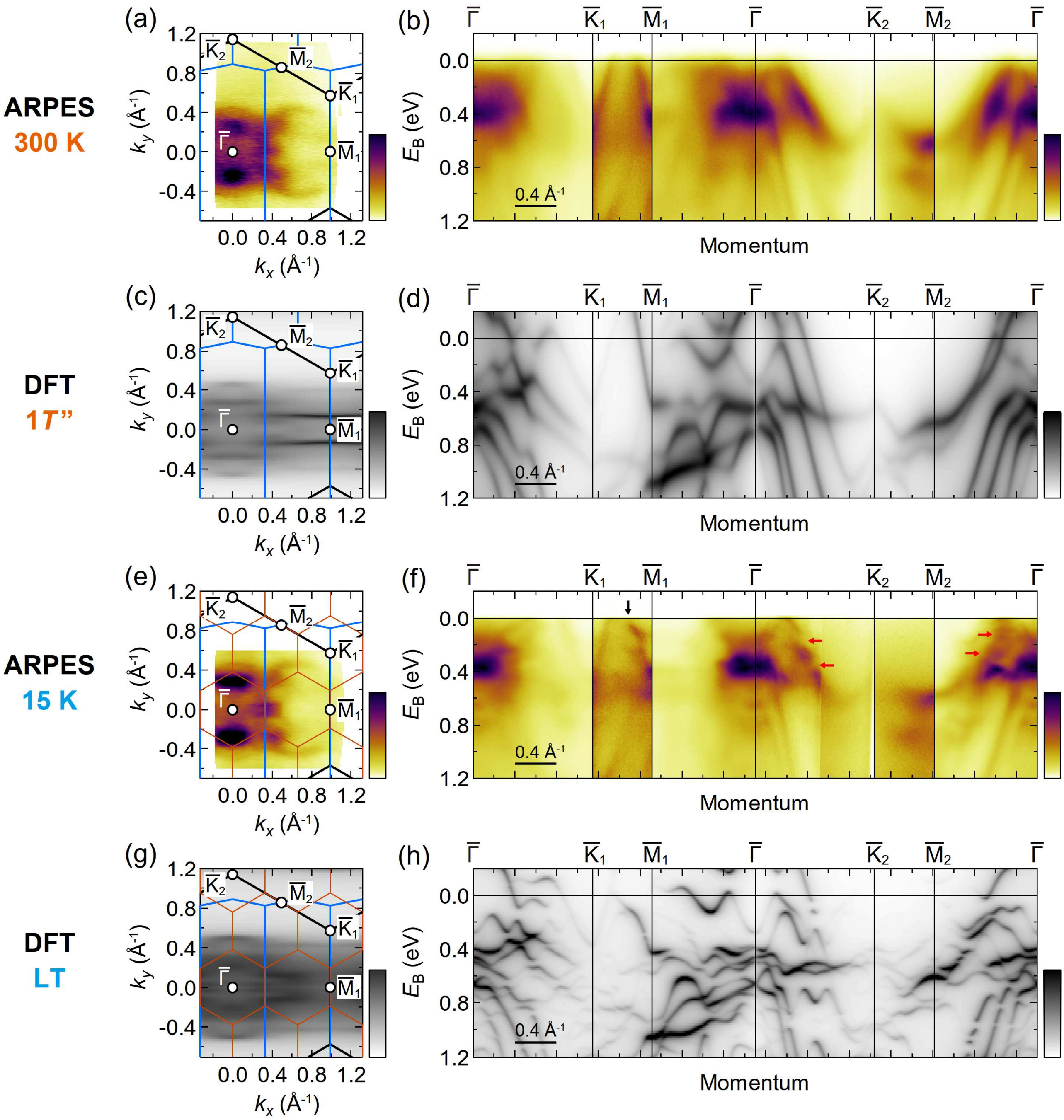

**Fig. 3. Overview of electronic band structure.** (a) ARPES intensity plot at $E_F$ in the 1*T*” phase (300 K, energy integral width: 20 meV) collected with a He-discharge lamp (photon energy: 21.2 eV). The regular and elongated hexagons indicate the (0 0 1) surface Brillouin zones of virtual 1*T* and 1*T*”, respectively. (b) ARPES spectra at 300 K along $\bar{\Gamma}-\bar{K}_{1(2)}-\bar{M}_{1(2)}-\bar{\Gamma}$. (c), (d) Band unfolding calculation for the 1*T*” phase ($k_z = 0$) of the Fermi surface [(c)] and band dispersions [(d)]. (e)–(h) Same as (a)–(d), but for the LT phase (ARPES data: 15 K). The orange hexagon indicates the (0 0 1) surface Brillouin zone of LT. The red arrows mark the characteristic spectral segmentation features. The black arrow highlights the conspicuous band modification appeared around $\bar{M}_1$.

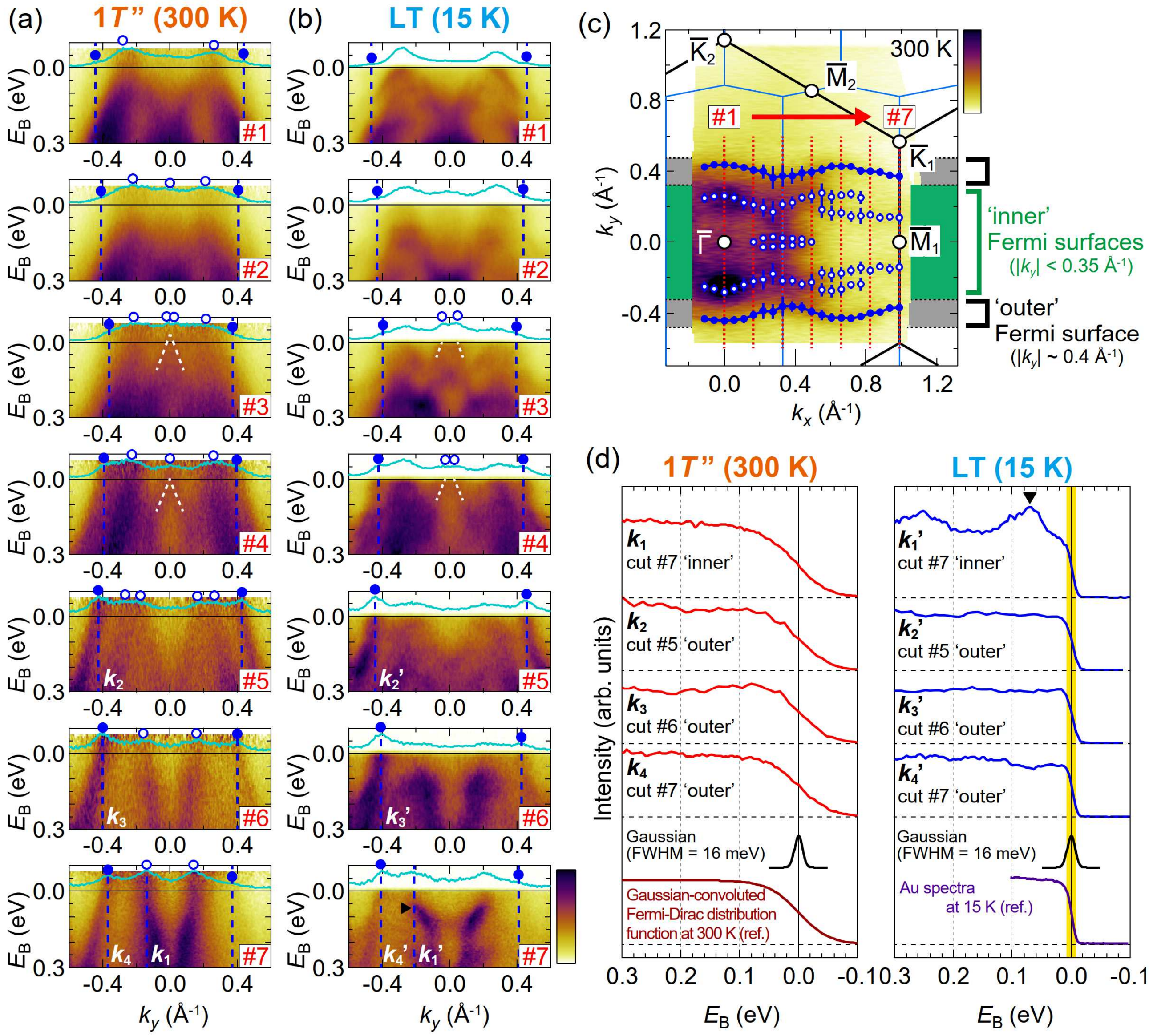

**Fig. 4. Temperature dependence of Fermi-level crossing bands.** (a), (b) ARPES spectra along selected momentums (cut #1–7, shown in (c)) for the 1*T*" (300 K) [(a)] and LT (15 K) phases [(b)]. The data of (a) is divided by the Fermi-Dirac distribution function to visualize the band structure near $E_F$. The cyan curves show the momentum distribution curves at $E_F$ (integral width: 20 meV). The blue filled circles (at $|k_y| \sim 0.4$ Å$^{-1}$) trace the 'outer' Fermi surface, whereas the open circles (in $|k_y| < 0.35$ Å$^{-1}$) partly track the complex 'inner' Fermi surfaces. The white broken lines in cut #3, 4 indicate the Λ-shaped band that forms the small Fermi pocket. The triangle marker in cut #7 at 15 K [(b)] depicts the location of the abrupt intensity suppression of the V-shaped band. (c) ARPES intensity map at $E_F$ at 300 K (same as Fig. 3(a)) overlaid with the peak plots extracted from the momentum distribution curves at $E_F$. (d), (e) Energy distribution curves (integral width: 0.05 Å$^{-1}$) at selected momentums [(i)–(iv) shown in (a), (b)] at 300 K [(d)] and 15 K [(e)]. The energy-resolution gaussian (FWHM of 16 meV), the simulated gaussian-convoluted Fermi-Dirac distribution function at 300 K, and the Au spectra at 15 K are also shown for reference.

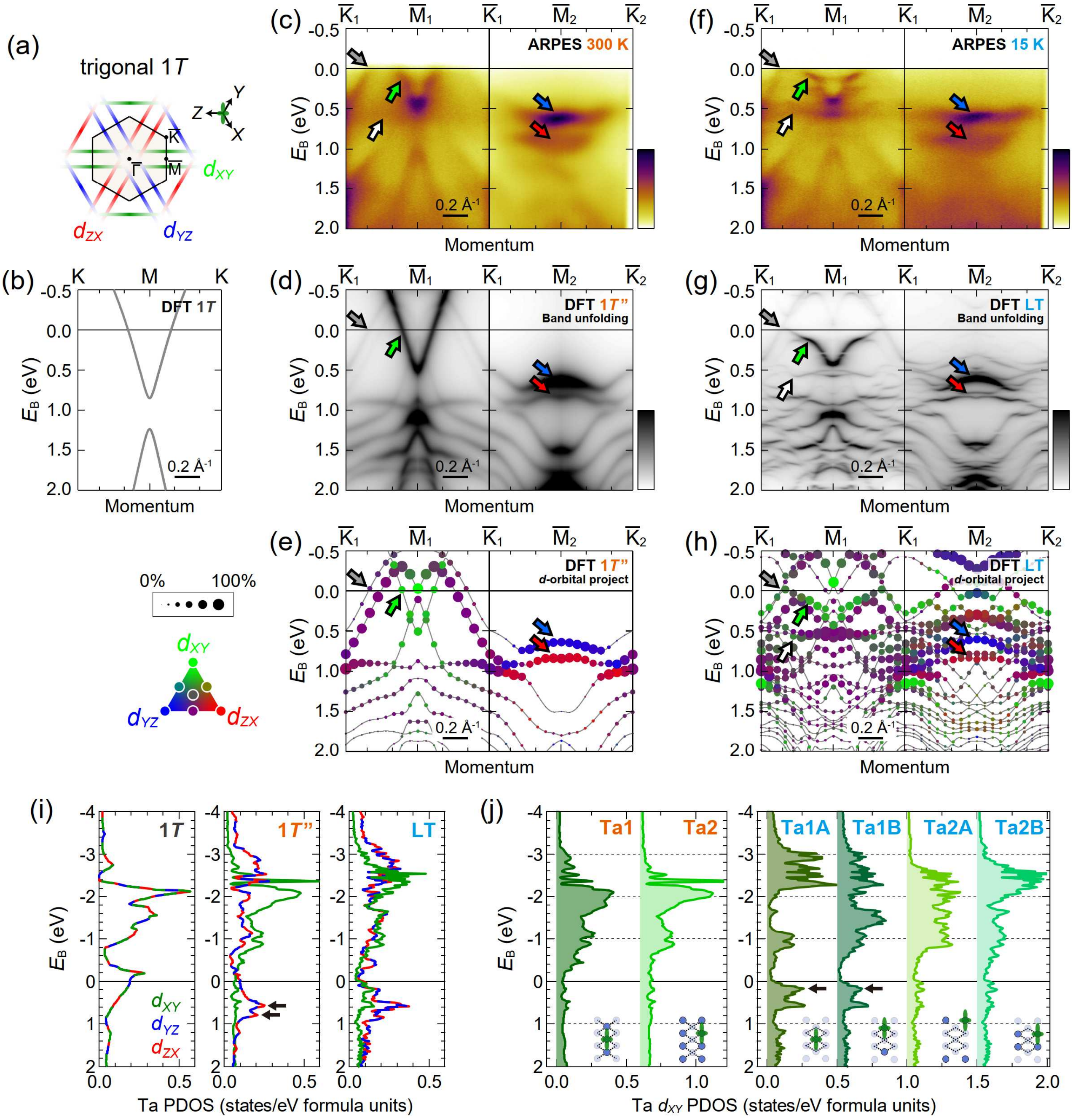

**Fig. 5. Orbital-dependent band reconstructions.** (a) Basic concept of the hidden Fermi surface proposed in the trigonal $1T$-$MX_2$. The three independent $t_{2g}$ ($d_{XY}/d_{YZ}/d_{ZX}$) σ-bonding form the Fermi surface around the Brillouin zone boundary. (b) Calculated band dispersion along K–M–K for the virtual $1T$-$TaTe_2$. (c)–(e) ARPES data (300 K) [(c)], band unfolding calculation [(d)], and Ta $t_{2g}$ orbital-projected band calculation [(e)] along $\overline{\mathrm{K}}_1-\overline{\mathrm{M}}_1-\overline{\mathrm{K}}_1$ and $\overline{\mathrm{K}}_1-\overline{\mathrm{M}}_2-\overline{\mathrm{K}}_2$ for the $1T$" phase. (f)–(h) Same as (c)–(e), but for the LT phase (ARPES data: 15 K). The green and red/blue arrows mark the $d_{XY}$-dominated V-shaped band and the $d_{YZ/ZX}$-dominated flat bands, respectively, whereas the gray arrow indicates the Te $p$-derived $E_F$-crossing band that forms the 'outer' Fermi surface. The while arrow depicts the band dispersion peculiar to the LT phase but its remnant is still observed in the ARPES spectra at 300 K [(e)]. (i) Calculated PDOS of the Ta site-averaged $d_{XY}/d_{YZ}/d_{ZX}$ for the virtual 1*T*, 1*T*'', and LT phases. The arrows mark the $d_{YZ}/d_{ZX}$ PDOS peaks in 1*T*'' corresponding to the flat bands around $\overline{\mathrm{M}}_2$. (j) Calculated PDOS of the Ta site-resolved $d_{XY}$ for the 1*T*'' and LT phases. The arrows indicate the $d_{YZ}/d_{ZX}$ PDOS peaks in Ta1A/1B sites (LT) that partly reflect the kink-like band structures appearing around $\overline{\mathrm{M}}_1$.